\documentclass[aps,prl,amsmath,amssymb,twocolumn,superscriptaddress]{revtex4-2}

\usepackage[dvipdfmx]{graphicx}
\usepackage{dcolumn}
\usepackage{bm}
\usepackage{here}
\usepackage{braket}
\usepackage{hyperref}
\hypersetup{
    colorlinks=true,
    linkcolor=blue,
    filecolor=blue,      
    urlcolor=blue,
}

\begin{document}

\title{Three-dimensional matter-wave interferometry of a trapped single ion}

\author{Ami Shinjo}
\affiliation{Graduate School of Engineering Science, Osaka University, 1-3 Machikaneyama, Toyonaka, Osaka 560-8531, Japan}
\author{Masato Baba}
\affiliation{Graduate School of Engineering Science, Osaka University, 1-3 Machikaneyama, Toyonaka, Osaka 560-8531, Japan}
\author{Koya Higashiyama}
\affiliation{Graduate School of Engineering Science, Osaka University, 1-3 Machikaneyama, Toyonaka, Osaka 560-8531, Japan}
\author{Ryoichi Saito}
\email{r\_saito@ee.es.osaka-u.ac.jp}
\affiliation{Graduate School of Engineering Science, Osaka University, 1-3 Machikaneyama, Toyonaka, Osaka 560-8531, Japan}
\affiliation{Quantum Information and Quantum Biology Division, Institute for Open and Transdisciplinary Research Initiatives,Osaka University, Osaka 560-8531, Japan.}

\author{Takashi Mukaiyama}
\email{muka@ee.es.osaka-u.ac.jp}
\affiliation{Graduate School of Engineering Science, Osaka University, 1-3 Machikaneyama, Toyonaka, Osaka 560-8531, Japan}
\affiliation{Quantum Information and Quantum Biology Division, Institute for Open and Transdisciplinary Research Initiatives,Osaka University, Osaka 560-8531, Japan.}




\date{\today}

\begin{abstract}
We report on a demonstration of Ramsey interferometry by three-dimensional motion with a trapped $^{171}$Yb$^+$ ion. We applied a momentum kick to the ion in a direction diagonal to the trap axes to initiate three-dimensional motion using a mode-locked pulse laser. The interference signal was analyzed theoretically to demonstrate three-dimensional matter-wave interference. This work paves the way to realizing matter-wave interferometry using trapped ions.
\end{abstract}


\maketitle


The wave nature of matter offers the potential of measuring physical quantities with high precision. Matter-wave interferometers normally exploit entanglement between the internal state and the motional state of matter. The accumulated matter-wave phases for different paths are derived from the interference signal of the internal states after the closing pulse of the interferometer. Since atoms and ions provide us with exquisite control of individual quanta using electronic, magnetic and optical means, they have been widely used as wave-like matters in many sensing applications. Neutral atom systems have been extensively used and there are a large number of demonstrations of precision sensing such as in measurement of the relativistic effects in electromagnetic interactions~\cite{Zeiske, Gorlitz, Yanagimachi}, atomic and molecular properties~\cite{Ekstrom1995, Schmiedmayer1995}, and measurement of inertial displacement~\cite{Kasevich, Oberthaler, Riele, Peters, Gustavson, Gustavson2000}.

In contrast, a quantum sensing based on a matter-wave interference with trapped ions has so far not been demonstrated. In an ion trap, the Schr\"{o}dinger cat state, which is the quantum superposition of classically distinct states, was realized for the first time for a laser-cooled ion prepared in the motional ground state~\cite{Monroe1996}. Later, a scheme to create the cat state using ultrashort pulses from a mode-locked laser was realized, and spin-motion entanglement was created on a short time scale~\cite{Campbell2010, Hayes, Mizrahi2013, Mizrahi2014}. This has also been achieved with ions in the thermal regime~\cite{Johnson2015, Johnson2017, Campos}. Campbell and Hamilton recently proposed a scheme to utilize a trapped ion in such a Schr\"{o}dinger cat state for a precise rotation sensing~\cite{campbell}. In their proposal, an ion undergoing two-dimensional circular motion is utilized to construct a Sagnac interferometer. Compared with neutral atom gyroscope, the ions with multiple circular motions make a large effective interferometer area with a compact physical size and also relatively immune to the change of acceleration. To realize the gyroscope with trapped ion in the future, interferometry for ions in motion along multiple symmetry axes is an important next step for the new types of sensing applications.

In this Letter, we demonstrate a matter-wave interferometry of a trapped ion in a three-dimensional motion, initiated by a momentum kick in a direction diagonal to any of the trap symmetry axes. We applied $\pi/2$ pulse to the ion, which puts it into a superposition state made up of the original spin state and motion, and the opposite spin state with additional momentum due to the momentum kick. This spin-motion coupling is applied along the three symmetry axes of the trap simultaneously, and half of the ion wave packet travels with a harmonic potential relative to the other half of the ion wave packet in the original spin and motion state. In the experiment, we applied second $\pi/2$ pulse after the interrogation time to close the interferometer. When the interrogation time of this Ramsey type interferometer is an integer multiple of the trap period, constructive interference is observed. Since trap frequencies for the three axes are different, the interference signal shows a complicated interrogation-time dependence. A slight change in the frequency of one of the trap axes causes a sensitive change in the interference signal, indicating that the interference arises from ion motion in three dimensions. We theoretically analyzed the interference signal and found good agreement between the measured data and the theory. We estimated the rotation sensing sensitivity assuming that the ion has a circular orbit with the current experimental parameters. Based on the evaluation, we mentioned our present system's proposal to the application of the gyroscope.


\begin{figure*}[tbp]
\centering
\includegraphics[scale=0.59]{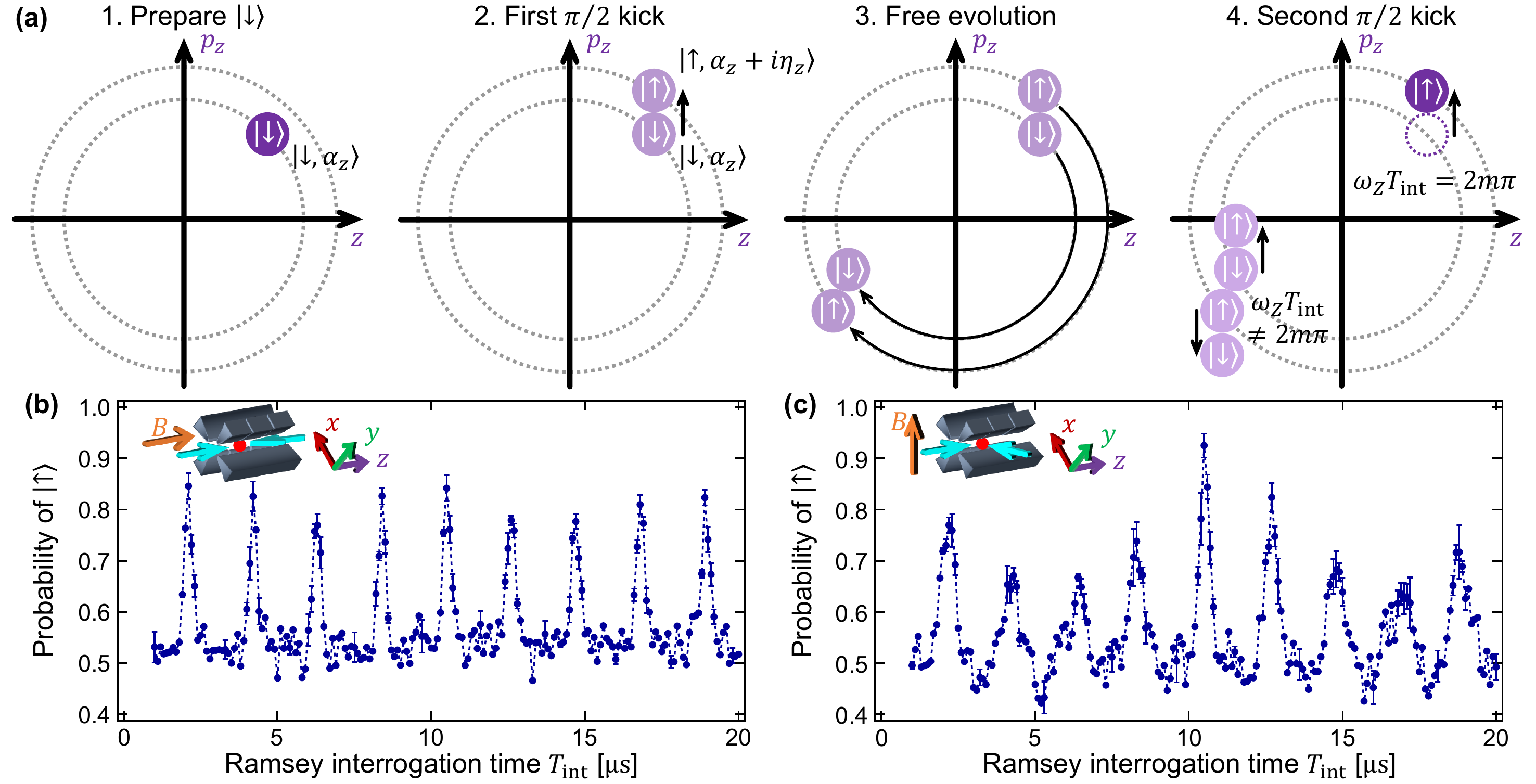}
\caption{
(a) Phase-space diagrams of one-dimensional interference. (b) Experimental configuration (shown in inset) and results for one-dimensional interference. The light blue arrows represent the direction of the pulse laser beam. The laser pulses have a linear-perpendicular-linear polarization configuration and their polarizations are both orthogonal to the quantization axis (defined by the magnetic field $B$). Each data point represents an average of 900 measurements. (c) Measured interference fringes for a diagonal momentum kick. 
Pulse trains are introduced orthogonally. 
Both pulse trains are horizontally polarized. Each data point represents an average of 600 measurements.
}
\label{fig:ramsey}
\end{figure*}

We trapped a single $^{171}$Yb$^+$ ion using a linear Paul trap. We used a mode-locked frequency-tripled Nd:YVO$_4$ laser with a pulse duration of $\tau \sim 15$ ps, a center wavelength of $2\pi/k = 355$ nm, and a repetition rate $f_{\mathrm{rep}} = 120.47$ MHz~\cite{Mizrahi2013, Mizrahi2014}. The laser drove stimulated Raman transitions between the two hyperfine ground states $\ket{\downarrow} = \ket{F=0, m_F=0 }$ and $\ket{\uparrow} = \ket{F=1, m_F=0}$ to give momentum kicks to the ion. Acousto-optic modulators (AOMs) were placed in both optical paths and were used to tune the frequency difference between the two beams to match the hyperfine splitting $\omega_{\mathrm{HF}} = n\omega_{\mathrm{rep}} + \omega_{\mathrm{AOM}}$, where $\omega_{\mathrm{AOM}}$ is the frequency shift generated by the AOMs. We chose $n=102$, and $\omega_{\mathrm{AOM}} =$ 354 MHz. Each optical path had the same length to arrive the pulse train at the ion simultaneously. All of 355 nm beam paths were covered with a case to prevent fluctuations of the beam spot position due to air flow. \par

By introducing the pulse trains from opposite directions along the axial trap axis shown in the inset of Fig.~\ref{fig:ramsey}(b), a one-dimensional momentum kick was given to the ion along the axial trap axis. Figure~\ref{fig:ramsey}(a) depicts the position-momentum phase space in the one-dimensional interference experiment. First, the ion was prepared in a $\ket{\downarrow}$ state using Doppler cooling and optical pumping. Therefore, the ion motional state is thermal, and is expressed as a superposition of coherent states. For simplicity, we express the initial state as a coherent state $\ket{\downarrow, \alpha_z}$ (step 1 in Fig.~\ref{fig:ramsey}(a)). Next, we applied a $\pi/2$ pulse using a 200 ns long pulse train. This pulse train is in the strong excitation regime~\cite{poyatos}, and applies the same momentum kick for each coherent state contained in the thermal state. The momentum kick $i\eta_z$ causes the ion to transfer to a superposition state $(\ket{\downarrow, \alpha_z} + \ket{\uparrow , \alpha_z+i \eta_z} ) / \sqrt{2}$ (step 2), where $\eta_z = \Delta k_z \sqrt{\hbar / 2M\omega_z}$ is the Lamb-Dicke parameter, $\Delta k_z$ is the difference in wave vector between the two beams, $M$ is the Ytterbium mass, and $\omega_z$ is the trap frequency of $z$ direction. After an interrogation time $T_{\mathrm{int}}$ (step 3), we applied a $\pi$/2 pulse again, which partially transfers the $\ket{\downarrow}$ state to a $\ket{\uparrow}$ state with added momentum $i\eta_z$, and transfers the $\ket{\uparrow}$ state to a $\ket{\downarrow}$ state with added momentum $-i\eta_z$ (step 4). Finally we detected the probability of the ion being in the $\ket{\uparrow}$ state. \par

\begin{figure*}[tbp]
\centering
\includegraphics[scale=0.63]{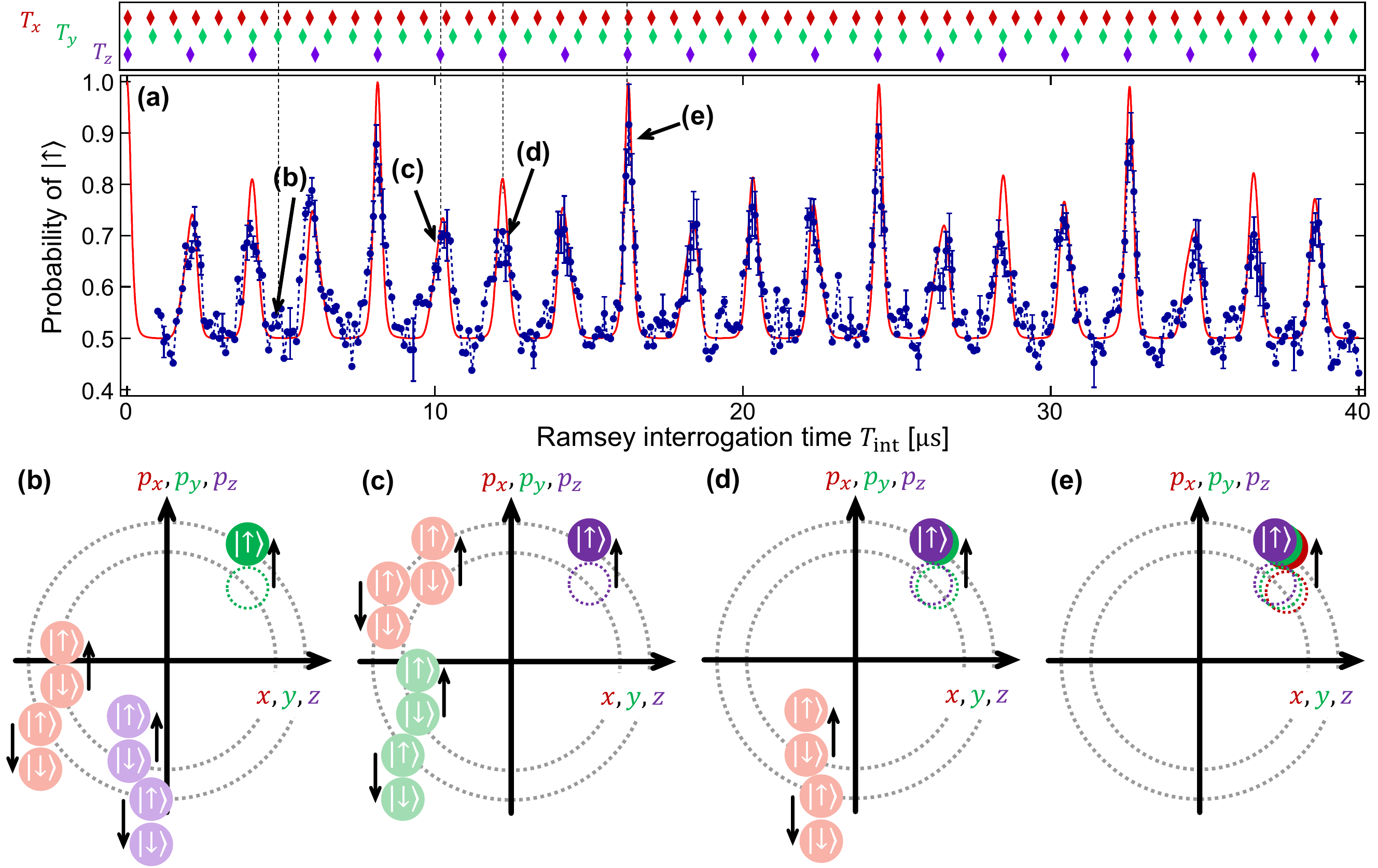}
\caption{Observed interference fringes at trap frequency $(\omega_x, \omega_y, \omega_z)= 2\pi \times(1229.0(8), 1349.7(7), 491.7(1))$ kHz. (a) Blue points show the experimental results (each data point represents an average of 900 measurements) and the red solid line is the fitted curve. (b--e) show phase-space diagrams at the timings indicated in (a). The red, green and purple circles represent the coherent state components in the $x$-direction, $y$-direction and $z$-direction, respectively. For simplicity, phase space trajectories along the $x$, $y$, and $z$ axes are shown in a single phase-space diagram without scaling.} \label{fig:ramseylong}
\end{figure*}

Figure~\ref{fig:ramsey}(b) shows the transition probability from state $\ket{\downarrow}$ to state $\ket{\uparrow}$ as a function of the interrogation time between two $\pi/2$ pulses for a one-dimensional momentum kick. As can be clearly seen, constructive interference appears at integer multiples of the trap period ($T_z$) determined by the axial trap frequency $\omega_z = 2\pi \times 475$ kHz. After the second $\pi$/2 pulse for $\omega_z T_{\mathrm{int}} = 2m\pi$ ($m$ is the integer in step 4 in Fig.~\ref{fig:ramsey}(a)), the transition probability reaches 0.8 when the two wave packets match, but is only around 0.5 when the wave packets did not match for $\omega_z T_{\mathrm{int}} \neq 2m\pi$. In Fig.~\ref{fig:ramsey}(b), the measured transition probability for the peaks is smaller than unity, because of imperfect Rabi oscillations due to the finite temperature.\par

Next, under the same trap conditions, we changed the direction of the momentum kick to along the direction diagonal to all trap axes. The results are shown in Fig.~\ref{fig:ramsey}(c). Peaks appear at integer multiples of the trap period, similar to the case for one-dimensional matter-wave interference in Fig.~\ref{fig:ramsey}(b). However, peak height variations from 0.6 to 0.9 are observed in this configuration, in contrast to the results shown in Fig.~\ref{fig:ramsey}(b); this is a clear indication of three-dimensional interference. \par

In order to clarify this complicated motion, we theoretically investigated three-dimensional matter-wave interference using a previously reported method~\cite{Mizrahi2014,mizrahithesis}. In our experimental scheme, we apply $\pi/2$ pulse to the ion. Therefore the pulse train operator becomes: 
\begin{eqnarray}
O_{N, \frac{\pi}{2}} = \frac{1}{\sqrt{2}} + \frac{i}{\sqrt{2}} \left( e^{i\phi_0} D[i\eta_x]D[i\eta_y]D[i\eta_z] \hat{\sigma}_+ \right. \nonumber \\ 
\left. + e^{-i\phi_0} D[-i\eta_x]D[-i\eta_y]D[-i\eta_z] \hat{\sigma}_- \right).
\label{eq:operator}
\end{eqnarray}
Here $D[i\eta_j] (j = x, y, z)$ is the harmonic oscillator displacement operator in phase space which moves coherent state to another coherent state with $\eta_j$ momentum transfer, and $\phi_0$ is an optical phase offset assumed it constant. After applying two $\pi/2$ pulses with the time interval $T_{\mathrm{int}}$ between pulse trains, the probability $P$ of detecting the ion in the $\ket{\uparrow}$ state after the interference process can be expressed as:
\begin{equation}
P = \frac{1}{2} + \frac{1}{2}P_x P_y P_z ,
\label{eq:ramsey}
\end{equation}
where $P_j (j=x, y, z)$ describes the wave function overlap along the $j$ axes:
\begin{equation}
P_j = e^{-(2\bar{n}_j + 1) \eta_j^2 (1- \cos (\omega_j T_{\mathrm{int}}))} \cos (\eta_j^2 \sin (\omega_j T_{\mathrm{int}})).
\label{eq:pj}
\end{equation}
Here $\bar{n}_j$ is the mean phonon number in the $j$ direction. 

To investigate the result of three-dimensional interferometry in detail, we observed the interference with a longer interrogation time. Figure~\ref{fig:ramseylong}(a) displays the results for interference initiated by the diagonal momentum kick. The red solid line shows the results of fitting using Eq.~(\ref{eq:ramsey}). The red, green, and purple diamonds at the top of each figure show the timing when integer multiples of $T_x$, $T_y$ and $T_z$, respectively, are obtained. Three trap frequencies $\omega_x, \omega_y, \omega_z$, and the ion temperature contained in $\bar{n}_j$ were taken as free parameters. From the fitting results, the trap frequencies were obtained to be $(\omega_x, \omega_y, \omega_z) = 2\pi \times (1229.0(8), 1349.7(7), 491.7(1))$ kHz and the ion temperature was 2.23(6) mK. On the other hand, we also measured radial trap frequency from the spectroscopy, the results were $(\omega_x, \omega_y) = 2\pi \times (1210(10), 1320(10))$ kHz, respectively. Also, we measured the calibration curve of axial trap frequency based on sideband spectroscopy, and the result corresponding to the condition of the measurement in Fig.~\ref{fig:ramseylong}(a) was $\omega_z = 2\pi \times 490(10)$ kHz. 
Hence, the fitting results of trap frequencies of three axes approximate to the measurement results using sideband spectroscopy and the calibration curve.
%

Figures~\ref{fig:ramseylong}(b--e) show phase-space diagrams for the ion for the timings shown in Fig.~\ref{fig:ramseylong}(a). At time (b), $T_{\mathrm{int}}$ is an integer multiple of the trap period along the $y$ axis ($T_y$), but not along the $x$ and $z$ axes ($T_x$, $T_z$). Constructive interference is not observed because the wave packets are in different positions along the $x$ and $z$ axes in the harmonic potential. 

When $T_{\mathrm{int}}$ is an integer multiple of $T_z$ (at time (c)), constructive interference is observed even though $T_{\mathrm{int}}$ is not an integer multiple of $T_x$ and $T_y$. The difference between times (b) and (c) arises from the fact that the trajectory along the $z$ axis is bigger than that along the $x$ and $y$ axes, due to the difference in the strength of the momentum kick and the trap confinement strength~\footnote{The momentum kick strength along $z$ axis is larger by a factor of $\sqrt{2}$ than that along the $x$ and $y$ axes, and the trap is shallower along the $z$ axis by a factor of 2.5 compared with the $x$ and $y$ axes.}.
At the time (e) when $T_{\mathrm{int}}$ is approximately commensurate with all three trap periods, the strongest peak appears due to perfect wave packet overlap. However, at the peaks the measured results were smaller than the calculated values, this is caused the imperfection of the spin flip due to the high ion temperature.

\begin{figure}[tbp]
\centering
\includegraphics[scale=0.5]{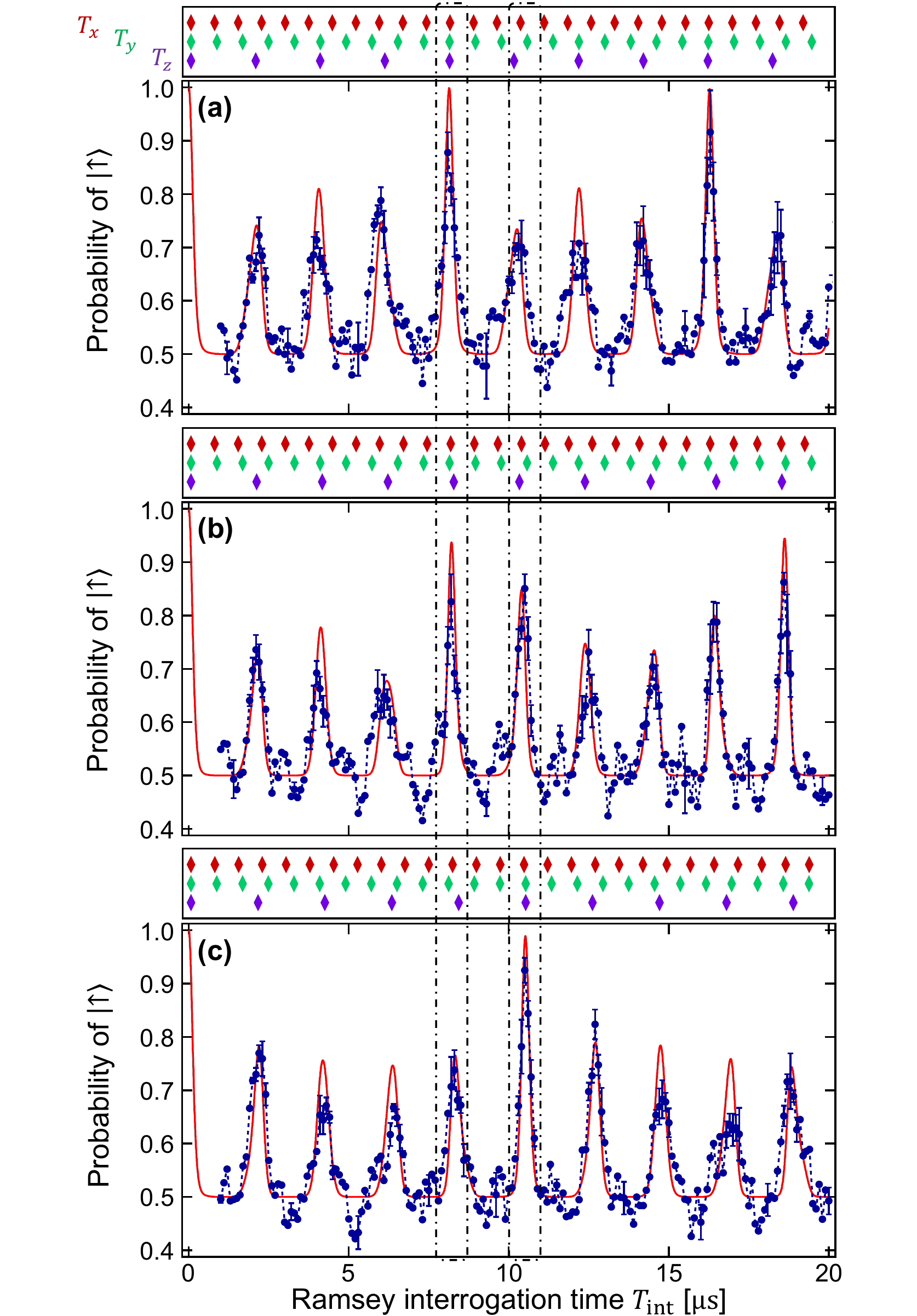}
\caption{Axial trap frequency dependence of three-dimensional interference. The axial trap frequency is shifted by changing the voltage applied to the ion trap.  Blue points represent an average of 900 or 600 measurements, and red solid lines are theoretical fits using Eq.~(\ref{eq:ramsey}). (a) $\omega_z = 2\pi \times 491.7(1)$ kHz. (b) $\omega_z = 2\pi \times 484.0(3)$ kHz. (c) $\omega_z = 2\pi \times 474.8(3)$ kHz.}
\label{fig:freq}
\end{figure}

To further investigate the three-dimensional interference, $\omega_z$ values of $2\pi \times$491.7(1) kHz, $2\pi \times$484.0(3) kHz, and $2\pi \times$474.8(3) kHz were used, as shown in Fig.~\ref{fig:freq}. The radial trap $\omega_x$ and $\omega_y$ values of each experiment obtained from the fitting were (a) $2\pi \times$1229.0(8) kHz and $2\pi \times$1349.7(9) kHz, (b) $2\pi \times$ 1229(2) kHz and $2\pi \times$1346(2) kHz, (c) $2\pi \times$1234(2) kHz and $2\pi \times$1335(2) kHz, respectively. Please pay attention to the peaks around 8.2 $\mu$s and 10.2 $\mu$s (the area surrounded by a dashed line). The peak near 8.2 $\mu$s decreases as $\omega_z$ increases. At $\omega_z = 2\pi \times 491.7$ kHz, a large interference signal is observed because the timing of the 4th period of $T_z$, the 11th period of $T_x$, and the 10th period of $T_y$, are matching. By changing $\omega_z$ to $2\pi \times 474.8$ kHz, the timing of the 4th period of $T_z$ does not match, and the interference signal strength is reduced.  In contrast, the peak near 10.2 $\mu$s increases as $\omega_z$ increases because the timing of the 5th period of $T_z$, the 14th period of $T_x$, and the 13th period of $T_y$ match at $\omega_z = 2\pi \times$ 474.8 kHz. 

\begin{figure}[tbp]
\centering
\includegraphics[scale=0.53]{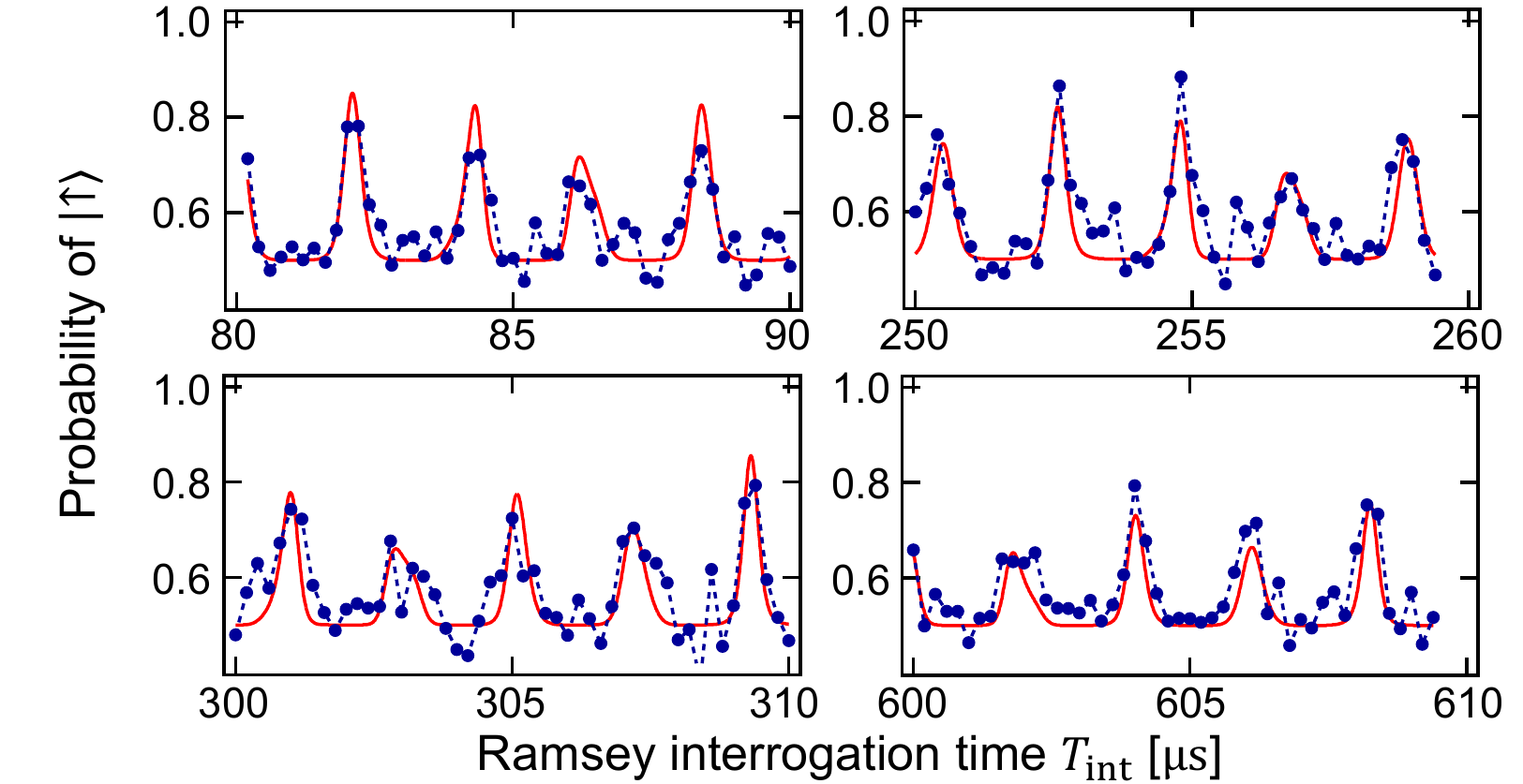}
\caption{Measured interference fringes for long interrogation times. Blue points represent an average of 300 measurements, and red solid lines are theoretical fits using Eq.~(\ref{eq:ramsey}) with an attenuation term $e^{-\gamma t}$.}
\label{fig:coherence}
\end{figure}

We next investigated the coherence time for three-dimensional ion motion. Figure~\ref{fig:coherence} shows the three-dimensional interference for various ranges of $T_{\mathrm{int}}$. In order to take into account the coherence time, we added an attenuation factor $e^{-\gamma t}$ into the second term of Eq.~(\ref{eq:ramsey}). Each red solid curve in Fig.~\ref{fig:coherence} is a best fit to the experimental results for each interrogation time for a fixed ion temperature and attenuation coefficient $\gamma$.  The results indicate that the decoherence time is 3.3(5) ms. Even for large $T_{\mathrm{int}}$ values up to 600 $\mu$s, the interference signal is well reproduced by the theoretical curves, indicating that the coherence of ion motion in all directions decays over the same time scale. 

In the current work, the multi-dimensional matter-wave interference was demonstrated in an anisotropic trapping potential, and therefore the ion was traveling in a Lissajous orbit. To apply the current system for rotation sensing, a circular ion motion needs to be realized by tuning the confinement strength of the trap to make the trap potential isotropic in two dimensions. Assuming that we realize an isotropic potential in the current trap condition, the expected sensitivity of the gyroscope is $S = 2 \times 10^5 \mathrm{\, deg}/\sqrt{\mathrm{hour}}$, which is eight orders of magnitude worse than the conventional ring laser gyroscope. Since the spatial separation of two wave packets in the current experiment is only on the order of 1 nm, which is the same scale with the zero point width, the sensitivity is quite limited. The sensitivity can be improved by adding much larger momentum kicks to the ion by increasing the number of pulses~\cite{Johnson2017}, reducing the trap confinement strength, and applying the displacement of the trap center~\cite{campbell}. With these contrivances, our system will attain the same accuracy as the ring laser gyroscope. Since the dynamics of the ion during the interrogation time reflects the imperfections of matter-wave interferometers~\cite{west}, the full understanding of the free time evolution of an ion is an important ingredient for improving the sensitivity of the matter-wave interferometers.

We acknowledge fruitful discussion with Mikio Kozuma and Ryotaro Inoue. This work was supported by JST-Mirai Program Grant Number JPMJMI17A3, Japan.

\bibliography{bunken}

\begin{thebibliography}{24}%
\makeatletter
\providecommand \@ifxundefined [1]{%
 \@ifx{#1\undefined}
}%
\providecommand \@ifnum [1]{%
 \ifnum #1\expandafter \@firstoftwo
 \else \expandafter \@secondoftwo
 \fi
}%
\providecommand \@ifx [1]{%
 \ifx #1\expandafter \@firstoftwo
 \else \expandafter \@secondoftwo
 \fi
}%
\providecommand \natexlab [1]{#1}%
\providecommand \enquote  [1]{``#1''}%
\providecommand \bibnamefont  [1]{#1}%
\providecommand \bibfnamefont [1]{#1}%
\providecommand \citenamefont [1]{#1}%
\providecommand \href@noop [0]{\@secondoftwo}%
\providecommand \href [0]{\begingroup \@sanitize@url \@href}%
\providecommand \@href[1]{\@@startlink{#1}\@@href}%
\providecommand \@@href[1]{\endgroup#1\@@endlink}%
\providecommand \@sanitize@url [0]{\catcode `\\12\catcode `\$12\catcode
  `\&12\catcode `\#12\catcode `\^12\catcode `\_12\catcode `\%12\relax}%
\providecommand \@@startlink[1]{}%
\providecommand \@@endlink[0]{}%
\providecommand \url  [0]{\begingroup\@sanitize@url \@url }%
\providecommand \@url [1]{\endgroup\@href {#1}{\urlprefix }}%
\providecommand \urlprefix  [0]{URL }%
\providecommand \Eprint [0]{\href }%
\providecommand \doibase [0]{https://doi.org/}%
\providecommand \selectlanguage [0]{\@gobble}%
\providecommand \bibinfo  [0]{\@secondoftwo}%
\providecommand \bibfield  [0]{\@secondoftwo}%
\providecommand \translation [1]{[#1]}%
\providecommand \BibitemOpen [0]{}%
\providecommand \bibitemStop [0]{}%
\providecommand \bibitemNoStop [0]{.\EOS\space}%
\providecommand \EOS [0]{\spacefactor3000\relax}%
\providecommand \BibitemShut  [1]{\csname bibitem#1\endcsname}%
\let\auto@bib@innerbib\@empty
\bibitem [{\citenamefont {Zeiske}\ \emph {et~al.}(1995)\citenamefont {Zeiske},
  \citenamefont {Zinner}, \citenamefont {Riehle},\ and\ \citenamefont
  {Helmcke}}]{Zeiske}%
  \BibitemOpen
  \bibfield  {author} {\bibinfo {author} {\bibfnamefont {K.}~\bibnamefont
  {Zeiske}}, \bibinfo {author} {\bibfnamefont {G.}~\bibnamefont {Zinner}},
  \bibinfo {author} {\bibfnamefont {F.}~\bibnamefont {Riehle}},\ and\ \bibinfo
  {author} {\bibfnamefont {J.}~\bibnamefont {Helmcke}},\ }\href
  {https://doi.org/10.1007/BF01135864} {\bibfield  {journal} {\bibinfo
  {journal} {Applied Physics B}\ }\textbf {\bibinfo {volume} {60}},\ \bibinfo
  {pages} {205} (\bibinfo {year} {1995})}\BibitemShut {NoStop}%
\bibitem [{\citenamefont {G\"orlitz}\ \emph {et~al.}(1995)\citenamefont
  {G\"orlitz}, \citenamefont {Schuh},\ and\ \citenamefont {Weis}}]{Gorlitz}%
  \BibitemOpen
  \bibfield  {author} {\bibinfo {author} {\bibfnamefont {A.}~\bibnamefont
  {G\"orlitz}}, \bibinfo {author} {\bibfnamefont {B.}~\bibnamefont {Schuh}},\
  and\ \bibinfo {author} {\bibfnamefont {A.}~\bibnamefont {Weis}},\ }\href
  {https://doi.org/10.1103/PhysRevA.51.R4305} {\bibfield  {journal} {\bibinfo
  {journal} {Phys. Rev. A}\ }\textbf {\bibinfo {volume} {51}},\ \bibinfo
  {pages} {R4305} (\bibinfo {year} {1995})}\BibitemShut {NoStop}%
\bibitem [{\citenamefont {Yanagimachi}\ \emph {et~al.}(2002)\citenamefont
  {Yanagimachi}, \citenamefont {Kajiro}, \citenamefont {Machiya},\ and\
  \citenamefont {Morinaga}}]{Yanagimachi}%
  \BibitemOpen
  \bibfield  {author} {\bibinfo {author} {\bibfnamefont {S.}~\bibnamefont
  {Yanagimachi}}, \bibinfo {author} {\bibfnamefont {M.}~\bibnamefont {Kajiro}},
  \bibinfo {author} {\bibfnamefont {M.}~\bibnamefont {Machiya}},\ and\ \bibinfo
  {author} {\bibfnamefont {A.}~\bibnamefont {Morinaga}},\ }\href
  {https://doi.org/10.1103/PhysRevA.65.042104} {\bibfield  {journal} {\bibinfo
  {journal} {Phys. Rev. A}\ }\textbf {\bibinfo {volume} {65}},\ \bibinfo
  {pages} {042104} (\bibinfo {year} {2002})}\BibitemShut {NoStop}%
\bibitem [{\citenamefont {Ekstrom}\ \emph {et~al.}(1995)\citenamefont
  {Ekstrom}, \citenamefont {Schmiedmayer}, \citenamefont {Chapman},
  \citenamefont {Hammond},\ and\ \citenamefont {Pritchard}}]{Ekstrom1995}%
  \BibitemOpen
  \bibfield  {author} {\bibinfo {author} {\bibfnamefont {C.~R.}\ \bibnamefont
  {Ekstrom}}, \bibinfo {author} {\bibfnamefont {J.}~\bibnamefont
  {Schmiedmayer}}, \bibinfo {author} {\bibfnamefont {M.~S.}\ \bibnamefont
  {Chapman}}, \bibinfo {author} {\bibfnamefont {T.~D.}\ \bibnamefont
  {Hammond}},\ and\ \bibinfo {author} {\bibfnamefont {D.~E.}\ \bibnamefont
  {Pritchard}},\ }\href {https://doi.org/10.1103/PhysRevA.51.3883} {\bibfield
  {journal} {\bibinfo  {journal} {Phys. Rev. A}\ }\textbf {\bibinfo {volume}
  {51}},\ \bibinfo {pages} {3883} (\bibinfo {year} {1995})}\BibitemShut
  {NoStop}%
\bibitem [{\citenamefont {Schmiedmayer}\ \emph {et~al.}(1995)\citenamefont
  {Schmiedmayer}, \citenamefont {Chapman}, \citenamefont {Ekstrom},
  \citenamefont {Hammond}, \citenamefont {Wehinger},\ and\ \citenamefont
  {Pritchard}}]{Schmiedmayer1995}%
  \BibitemOpen
  \bibfield  {author} {\bibinfo {author} {\bibfnamefont {J.}~\bibnamefont
  {Schmiedmayer}}, \bibinfo {author} {\bibfnamefont {M.~S.}\ \bibnamefont
  {Chapman}}, \bibinfo {author} {\bibfnamefont {C.~R.}\ \bibnamefont
  {Ekstrom}}, \bibinfo {author} {\bibfnamefont {T.~D.}\ \bibnamefont
  {Hammond}}, \bibinfo {author} {\bibfnamefont {S.}~\bibnamefont {Wehinger}},\
  and\ \bibinfo {author} {\bibfnamefont {D.~E.}\ \bibnamefont {Pritchard}},\
  }\href {https://doi.org/10.1103/PhysRevLett.74.1043} {\bibfield  {journal}
  {\bibinfo  {journal} {Phys. Rev. Lett.}\ }\textbf {\bibinfo {volume} {74}},\
  \bibinfo {pages} {1043} (\bibinfo {year} {1995})}\BibitemShut {NoStop}%
\bibitem [{\citenamefont {Kasevich}\ and\ \citenamefont
  {Chu}(1991)}]{Kasevich}%
  \BibitemOpen
  \bibfield  {author} {\bibinfo {author} {\bibfnamefont {M.}~\bibnamefont
  {Kasevich}}\ and\ \bibinfo {author} {\bibfnamefont {S.}~\bibnamefont {Chu}},\
  }\href {https://doi.org/10.1103/PhysRevLett.67.181} {\bibfield  {journal}
  {\bibinfo  {journal} {Phys. Rev. Lett.}\ }\textbf {\bibinfo {volume} {67}},\
  \bibinfo {pages} {181} (\bibinfo {year} {1991})}\BibitemShut {NoStop}%
\bibitem [{\citenamefont {Oberthaler}\ \emph {et~al.}(1996)\citenamefont
  {Oberthaler}, \citenamefont {Bernet}, \citenamefont {Rasel}, \citenamefont
  {Schmiedmayer},\ and\ \citenamefont {Zeilinger}}]{Oberthaler}%
  \BibitemOpen
  \bibfield  {author} {\bibinfo {author} {\bibfnamefont {M.~K.}\ \bibnamefont
  {Oberthaler}}, \bibinfo {author} {\bibfnamefont {S.}~\bibnamefont {Bernet}},
  \bibinfo {author} {\bibfnamefont {E.~M.}\ \bibnamefont {Rasel}}, \bibinfo
  {author} {\bibfnamefont {J.}~\bibnamefont {Schmiedmayer}},\ and\ \bibinfo
  {author} {\bibfnamefont {A.}~\bibnamefont {Zeilinger}},\ }\href
  {https://doi.org/10.1103/PhysRevA.54.3165} {\bibfield  {journal} {\bibinfo
  {journal} {Phys. Rev. A}\ }\textbf {\bibinfo {volume} {54}},\ \bibinfo
  {pages} {3165} (\bibinfo {year} {1996})}\BibitemShut {NoStop}%
\bibitem [{\citenamefont {Riehle}\ \emph {et~al.}(1991)\citenamefont {Riehle},
  \citenamefont {Kisters}, \citenamefont {Witte}, \citenamefont {Helmcke},\
  and\ \citenamefont {Bord\'e}}]{Riele}%
  \BibitemOpen
  \bibfield  {author} {\bibinfo {author} {\bibfnamefont {F.}~\bibnamefont
  {Riehle}}, \bibinfo {author} {\bibfnamefont {T.}~\bibnamefont {Kisters}},
  \bibinfo {author} {\bibfnamefont {A.}~\bibnamefont {Witte}}, \bibinfo
  {author} {\bibfnamefont {J.}~\bibnamefont {Helmcke}},\ and\ \bibinfo {author}
  {\bibfnamefont {C.~J.}\ \bibnamefont {Bord\'e}},\ }\href
  {https://doi.org/10.1103/PhysRevLett.67.177} {\bibfield  {journal} {\bibinfo
  {journal} {Phys. Rev. Lett.}\ }\textbf {\bibinfo {volume} {67}},\ \bibinfo
  {pages} {177} (\bibinfo {year} {1991})}\BibitemShut {NoStop}%
\bibitem [{\citenamefont {Peters}\ \emph {et~al.}(1999)\citenamefont {Peters},
  \citenamefont {Chung},\ and\ \citenamefont {Chu}}]{Peters}%
  \BibitemOpen
  \bibfield  {author} {\bibinfo {author} {\bibfnamefont {A.}~\bibnamefont
  {Peters}}, \bibinfo {author} {\bibfnamefont {K.~Y.}\ \bibnamefont {Chung}},\
  and\ \bibinfo {author} {\bibfnamefont {S.}~\bibnamefont {Chu}},\ }\href
  {https://doi.org/10.1038/23655} {\bibfield  {journal} {\bibinfo  {journal}
  {Nature}\ }\textbf {\bibinfo {volume} {400}},\ \bibinfo {pages} {849}
  (\bibinfo {year} {1999})}\BibitemShut {NoStop}%
\bibitem [{\citenamefont {Gustavson}\ \emph {et~al.}(1997)\citenamefont
  {Gustavson}, \citenamefont {Bouyer},\ and\ \citenamefont
  {Kasevich}}]{Gustavson}%
  \BibitemOpen
  \bibfield  {author} {\bibinfo {author} {\bibfnamefont {T.~L.}\ \bibnamefont
  {Gustavson}}, \bibinfo {author} {\bibfnamefont {P.}~\bibnamefont {Bouyer}},\
  and\ \bibinfo {author} {\bibfnamefont {M.~A.}\ \bibnamefont {Kasevich}},\
  }\href {https://doi.org/10.1103/PhysRevLett.78.2046} {\bibfield  {journal}
  {\bibinfo  {journal} {Phys. Rev. Lett.}\ }\textbf {\bibinfo {volume} {78}},\
  \bibinfo {pages} {2046} (\bibinfo {year} {1997})}\BibitemShut {NoStop}%
\bibitem [{\citenamefont {Gustavson}\ \emph {et~al.}(2000)\citenamefont
  {Gustavson}, \citenamefont {Landragin},\ and\ \citenamefont
  {Kasevich}}]{Gustavson2000}%
  \BibitemOpen
  \bibfield  {author} {\bibinfo {author} {\bibfnamefont {T.~L.}\ \bibnamefont
  {Gustavson}}, \bibinfo {author} {\bibfnamefont {A.}~\bibnamefont
  {Landragin}},\ and\ \bibinfo {author} {\bibfnamefont {M.~A.}\ \bibnamefont
  {Kasevich}},\ }\href {https://doi.org/10.1088/0264-9381/17/12/311} {\bibfield
   {journal} {\bibinfo  {journal} {Classical and Quantum Gravity}\ }\textbf
  {\bibinfo {volume} {17}},\ \bibinfo {pages} {2385} (\bibinfo {year}
  {2000})}\BibitemShut {NoStop}%
\bibitem [{\citenamefont {Monroe}\ \emph {et~al.}(1996)\citenamefont {Monroe},
  \citenamefont {Meekhof}, \citenamefont {King},\ and\ \citenamefont
  {Wineland}}]{Monroe1996}%
  \BibitemOpen
  \bibfield  {author} {\bibinfo {author} {\bibfnamefont {C.}~\bibnamefont
  {Monroe}}, \bibinfo {author} {\bibfnamefont {D.~M.}\ \bibnamefont {Meekhof}},
  \bibinfo {author} {\bibfnamefont {B.~E.}\ \bibnamefont {King}},\ and\
  \bibinfo {author} {\bibfnamefont {D.~J.}\ \bibnamefont {Wineland}},\ }\href
  {https://doi.org/10.1126/science.272.5265.1131} {\bibfield  {journal}
  {\bibinfo  {journal} {Science}\ }\textbf {\bibinfo {volume} {272}},\ \bibinfo
  {pages} {1131} (\bibinfo {year} {1996})}\BibitemShut {NoStop}%
\bibitem [{\citenamefont {Campbell}\ \emph {et~al.}(2010)\citenamefont
  {Campbell}, \citenamefont {Mizrahi}, \citenamefont {Quraishi}, \citenamefont
  {Senko}, \citenamefont {Hayes}, \citenamefont {Hucul}, \citenamefont
  {Matsukevich}, \citenamefont {Maunz},\ and\ \citenamefont
  {Monroe}}]{Campbell2010}%
  \BibitemOpen
  \bibfield  {author} {\bibinfo {author} {\bibfnamefont {W.~C.}\ \bibnamefont
  {Campbell}}, \bibinfo {author} {\bibfnamefont {J.}~\bibnamefont {Mizrahi}},
  \bibinfo {author} {\bibfnamefont {Q.}~\bibnamefont {Quraishi}}, \bibinfo
  {author} {\bibfnamefont {C.}~\bibnamefont {Senko}}, \bibinfo {author}
  {\bibfnamefont {D.}~\bibnamefont {Hayes}}, \bibinfo {author} {\bibfnamefont
  {D.}~\bibnamefont {Hucul}}, \bibinfo {author} {\bibfnamefont {D.~N.}\
  \bibnamefont {Matsukevich}}, \bibinfo {author} {\bibfnamefont
  {P.}~\bibnamefont {Maunz}},\ and\ \bibinfo {author} {\bibfnamefont
  {C.}~\bibnamefont {Monroe}},\ }\href
  {https://doi.org/10.1103/PhysRevLett.105.090502} {\bibfield  {journal}
  {\bibinfo  {journal} {Phys. Rev. Lett.}\ }\textbf {\bibinfo {volume} {105}},\
  \bibinfo {pages} {090502} (\bibinfo {year} {2010})}\BibitemShut {NoStop}%
\bibitem [{\citenamefont {Hayes}\ \emph {et~al.}(2010)\citenamefont {Hayes},
  \citenamefont {Matsukevich}, \citenamefont {Maunz}, \citenamefont {Hucul},
  \citenamefont {Quraishi}, \citenamefont {Olmschenk}, \citenamefont
  {Campbell}, \citenamefont {Mizrahi}, \citenamefont {Senko},\ and\
  \citenamefont {Monroe}}]{Hayes}%
  \BibitemOpen
  \bibfield  {author} {\bibinfo {author} {\bibfnamefont {D.}~\bibnamefont
  {Hayes}}, \bibinfo {author} {\bibfnamefont {D.~N.}\ \bibnamefont
  {Matsukevich}}, \bibinfo {author} {\bibfnamefont {P.}~\bibnamefont {Maunz}},
  \bibinfo {author} {\bibfnamefont {D.}~\bibnamefont {Hucul}}, \bibinfo
  {author} {\bibfnamefont {Q.}~\bibnamefont {Quraishi}}, \bibinfo {author}
  {\bibfnamefont {S.}~\bibnamefont {Olmschenk}}, \bibinfo {author}
  {\bibfnamefont {W.}~\bibnamefont {Campbell}}, \bibinfo {author}
  {\bibfnamefont {J.}~\bibnamefont {Mizrahi}}, \bibinfo {author} {\bibfnamefont
  {C.}~\bibnamefont {Senko}},\ and\ \bibinfo {author} {\bibfnamefont
  {C.}~\bibnamefont {Monroe}},\ }\href
  {https://doi.org/10.1103/PhysRevLett.104.140501} {\bibfield  {journal}
  {\bibinfo  {journal} {Phys. Rev. Lett.}\ }\textbf {\bibinfo {volume} {104}},\
  \bibinfo {pages} {140501} (\bibinfo {year} {2010})}\BibitemShut {NoStop}%
\bibitem [{\citenamefont {Mizrahi}\ \emph {et~al.}(2013)\citenamefont
  {Mizrahi}, \citenamefont {Senko}, \citenamefont {Neyenhuis}, \citenamefont
  {Johnson}, \citenamefont {Campbell}, \citenamefont {Conover},\ and\
  \citenamefont {Monroe}}]{Mizrahi2013}%
  \BibitemOpen
  \bibfield  {author} {\bibinfo {author} {\bibfnamefont {J.}~\bibnamefont
  {Mizrahi}}, \bibinfo {author} {\bibfnamefont {C.}~\bibnamefont {Senko}},
  \bibinfo {author} {\bibfnamefont {B.}~\bibnamefont {Neyenhuis}}, \bibinfo
  {author} {\bibfnamefont {K.~G.}\ \bibnamefont {Johnson}}, \bibinfo {author}
  {\bibfnamefont {W.~C.}\ \bibnamefont {Campbell}}, \bibinfo {author}
  {\bibfnamefont {C.~W.~S.}\ \bibnamefont {Conover}},\ and\ \bibinfo {author}
  {\bibfnamefont {C.}~\bibnamefont {Monroe}},\ }\href
  {https://doi.org/10.1103/PhysRevLett.110.203001} {\bibfield  {journal}
  {\bibinfo  {journal} {Phys. Rev. Lett.}\ }\textbf {\bibinfo {volume} {110}},\
  \bibinfo {pages} {203001} (\bibinfo {year} {2013})}\BibitemShut {NoStop}%
\bibitem [{\citenamefont {Mizrahi}\ \emph {et~al.}(2014)\citenamefont
  {Mizrahi}, \citenamefont {Neyenhuis}, \citenamefont {Johnson}, \citenamefont
  {Campbell}, \citenamefont {Senko}, \citenamefont {Hayes},\ and\ \citenamefont
  {Monroe}}]{Mizrahi2014}%
  \BibitemOpen
  \bibfield  {author} {\bibinfo {author} {\bibfnamefont {J.}~\bibnamefont
  {Mizrahi}}, \bibinfo {author} {\bibfnamefont {B.}~\bibnamefont {Neyenhuis}},
  \bibinfo {author} {\bibfnamefont {K.~G.}\ \bibnamefont {Johnson}}, \bibinfo
  {author} {\bibfnamefont {W.~C.}\ \bibnamefont {Campbell}}, \bibinfo {author}
  {\bibfnamefont {C.}~\bibnamefont {Senko}}, \bibinfo {author} {\bibfnamefont
  {D.}~\bibnamefont {Hayes}},\ and\ \bibinfo {author} {\bibfnamefont
  {C.}~\bibnamefont {Monroe}},\ }\href
  {https://doi.org/10.1007/s00340-013-5717-6} {\bibfield  {journal} {\bibinfo
  {journal} {Applied Physics B}\ }\textbf {\bibinfo {volume} {114}},\ \bibinfo
  {pages} {45} (\bibinfo {year} {2014})}\BibitemShut {NoStop}%
\bibitem [{\citenamefont {Johnson}\ \emph {et~al.}(2015)\citenamefont
  {Johnson}, \citenamefont {Neyenhuis}, \citenamefont {Mizrahi}, \citenamefont
  {Wong-Campos},\ and\ \citenamefont {Monroe}}]{Johnson2015}%
  \BibitemOpen
  \bibfield  {author} {\bibinfo {author} {\bibfnamefont {K.~G.}\ \bibnamefont
  {Johnson}}, \bibinfo {author} {\bibfnamefont {B.}~\bibnamefont {Neyenhuis}},
  \bibinfo {author} {\bibfnamefont {J.}~\bibnamefont {Mizrahi}}, \bibinfo
  {author} {\bibfnamefont {J.~D.}\ \bibnamefont {Wong-Campos}},\ and\ \bibinfo
  {author} {\bibfnamefont {C.}~\bibnamefont {Monroe}},\ }\href
  {https://doi.org/10.1103/PhysRevLett.115.213001} {\bibfield  {journal}
  {\bibinfo  {journal} {Phys. Rev. Lett.}\ }\textbf {\bibinfo {volume} {115}},\
  \bibinfo {pages} {213001} (\bibinfo {year} {2015})}\BibitemShut {NoStop}%
\bibitem [{\citenamefont {Johnson}\ \emph {et~al.}(2017)\citenamefont
  {Johnson}, \citenamefont {Wong-Campos}, \citenamefont {Neyenhuis},
  \citenamefont {Mizrahi},\ and\ \citenamefont {Monroe}}]{Johnson2017}%
  \BibitemOpen
  \bibfield  {author} {\bibinfo {author} {\bibfnamefont {K.~G.}\ \bibnamefont
  {Johnson}}, \bibinfo {author} {\bibfnamefont {J.~D.}\ \bibnamefont
  {Wong-Campos}}, \bibinfo {author} {\bibfnamefont {B.}~\bibnamefont
  {Neyenhuis}}, \bibinfo {author} {\bibfnamefont {J.}~\bibnamefont {Mizrahi}},\
  and\ \bibinfo {author} {\bibfnamefont {C.}~\bibnamefont {Monroe}},\ }\href
  {https://doi.org/10.1038/s41467-017-00682-6} {\bibfield  {journal} {\bibinfo
  {journal} {Nature Communications}\ }\textbf {\bibinfo {volume} {8}},\
  \bibinfo {pages} {697} (\bibinfo {year} {2017})}\BibitemShut {NoStop}%
\bibitem [{\citenamefont {Wong-Campos}\ \emph {et~al.}(2017)\citenamefont
  {Wong-Campos}, \citenamefont {Moses}, \citenamefont {Johnson},\ and\
  \citenamefont {Monroe}}]{Campos}%
  \BibitemOpen
  \bibfield  {author} {\bibinfo {author} {\bibfnamefont {J.~D.}\ \bibnamefont
  {Wong-Campos}}, \bibinfo {author} {\bibfnamefont {S.~A.}\ \bibnamefont
  {Moses}}, \bibinfo {author} {\bibfnamefont {K.~G.}\ \bibnamefont {Johnson}},\
  and\ \bibinfo {author} {\bibfnamefont {C.}~\bibnamefont {Monroe}},\ }\href
  {https://doi.org/10.1103/PhysRevLett.119.230501} {\bibfield  {journal}
  {\bibinfo  {journal} {Phys. Rev. Lett.}\ }\textbf {\bibinfo {volume} {119}},\
  \bibinfo {pages} {230501} (\bibinfo {year} {2017})}\BibitemShut {NoStop}%
\bibitem [{\citenamefont {Campbell}\ and\ \citenamefont
  {Hamilton}(2017)}]{campbell}%
  \BibitemOpen
  \bibfield  {author} {\bibinfo {author} {\bibfnamefont {W.~C.}\ \bibnamefont
  {Campbell}}\ and\ \bibinfo {author} {\bibfnamefont {P.}~\bibnamefont
  {Hamilton}},\ }\href {https://doi.org/10.1088/1361-6455/aa5a8f} {\bibfield
  {journal} {\bibinfo  {journal} {J. Phys. B: At. Mol. Opt. Phys.}\ }\textbf
  {\bibinfo {volume} {50}},\ \bibinfo {pages} {064002} (\bibinfo {year}
  {2017})}\BibitemShut {NoStop}%
\bibitem [{\citenamefont {Poyatos}\ \emph {et~al.}(1996)\citenamefont
  {Poyatos}, \citenamefont {Cirac}, \citenamefont {Blatt},\ and\ \citenamefont
  {Zoller}}]{poyatos}%
  \BibitemOpen
  \bibfield  {author} {\bibinfo {author} {\bibfnamefont {J.~F.}\ \bibnamefont
  {Poyatos}}, \bibinfo {author} {\bibfnamefont {J.~I.}\ \bibnamefont {Cirac}},
  \bibinfo {author} {\bibfnamefont {R.}~\bibnamefont {Blatt}},\ and\ \bibinfo
  {author} {\bibfnamefont {P.}~\bibnamefont {Zoller}},\ }\href
  {https://doi.org/10.1103/PhysRevA.54.1532} {\bibfield  {journal} {\bibinfo
  {journal} {Phys. Rev. A}\ }\textbf {\bibinfo {volume} {54}},\ \bibinfo
  {pages} {1532} (\bibinfo {year} {1996})}\BibitemShut {NoStop}%
\bibitem [{\citenamefont {Mizrahi}(2013)}]{mizrahithesis}%
  \BibitemOpen
  \bibfield  {author} {\bibinfo {author} {\bibfnamefont {J.~A.}\ \bibnamefont
  {Mizrahi}},\ }\emph {\bibinfo {title} {Ultrafast control of spin and motion
  in trapped ions}},\ \href@noop {} {Ph.D. thesis} (\bibinfo {year}
  {2013})\BibitemShut {NoStop}%
\bibitem [{Note1()}]{Note1}%
  \BibitemOpen
  \bibinfo {note} {The momentum kick strength along $z$ axis is larger by a
  factor of $\protect \sqrt {2}$ than that along the $x$ and $y$ axes, and the
  trap is shallower along the $z$ axis by a factor of 2.5 compared with the $x$
  and $y$ axes.}\BibitemShut {Stop}%
\bibitem [{\citenamefont {West}(2019)}]{west}%
  \BibitemOpen
  \bibfield  {author} {\bibinfo {author} {\bibfnamefont {A.~D.}\ \bibnamefont
  {West}},\ }\href {https://doi.org/10.1103/PhysRevA.100.063622} {\bibfield
  {journal} {\bibinfo  {journal} {Phys. Rev. A}\ }\textbf {\bibinfo {volume}
  {100}},\ \bibinfo {pages} {063622} (\bibinfo {year} {2019})}\BibitemShut
  {NoStop}%
\end{thebibliography}%
\end{document}



\title{Supplemental Material: Three-dimensional matter-wave interferometry of a trapped single ion}

\author{Ami Shinjo}
\affiliation{Graduate School of Engineering Science, Osaka University, 1-3 Machikaneyama, Toyonaka, Osaka 560-8531, Japan}
\author{Masato Baba}
\affiliation{Graduate School of Engineering Science, Osaka University, 1-3 Machikaneyama, Toyonaka, Osaka 560-8531, Japan}
\author{Koya Higashiyama}
\affiliation{Graduate School of Engineering Science, Osaka University, 1-3 Machikaneyama, Toyonaka, Osaka 560-8531, Japan}
\author{Ryoichi Saito}
\email{r\_saito@ee.es.osaka-u.ac.jp}
\affiliation{Graduate School of Engineering Science, Osaka University, 1-3 Machikaneyama, Toyonaka, Osaka 560-8531, Japan}
\affiliation{Quantum Information and Quantum Biology Division, Institute for Open and Transdisciplinary Research Initiatives,Osaka University, Osaka 560-8531, Japan.}

\author{Takashi Mukaiyama}
\email{muka@ee.es.osaka-u.ac.jp}
\affiliation{Graduate School of Engineering Science, Osaka University, 1-3 Machikaneyama, Toyonaka, Osaka 560-8531, Japan}
\affiliation{Quantum Information and Quantum Biology Division, Institute for Open and Transdisciplinary Research Initiatives,Osaka University, Osaka 560-8531, Japan.}




\date{\today}



\maketitle


\section{Derivation of Eq.~(1) and Eq.~(2) }
In our manuscript, we apply short Raman pulse trains which are in the fast regime called by Ref.~[22] to the initial state $\ket{\downarrow, \alpha_x, \alpha_y, \alpha_z} \equiv \ket{\Psi}$, the pulse train operator of our scheme can be defined like Eq.~(5.43) in Ref.~[22]:
\begin{equation}
O_N = \cos \frac{\Theta}{2} + i\sin \frac{\Theta}{2} \left( e^{i\phi_0} D[i\eta] \hat{\sigma}_+ + e^{-i\phi_0} D[-i\eta] \hat{\sigma}_- \right).
\label{eq:1}
\end{equation}
In our experiment, we irradiate two $\pi/2$ pulses, then apply $\Theta = \pi/2$ to Eq.~(\ref{eq:1}), we obtain
\begin{equation}
O_{N, \frac{\pi}{2}} = \frac{1}{\sqrt{2}} + \frac{i}{\sqrt{2}} \left( e^{i\phi_0} D[i\eta] \hat{\sigma}_+ + e^{-i\phi_0} D[-i\eta] \hat{\sigma}_- \right). \nonumber
\label{eq:2}
\end{equation}
Since we give momentum kick in three-dimensional direction to the ion, the pulse train operator should be expanded in three-dimension:
\begin{equation}
O_{N, \frac{\pi}{2}} = \frac{1}{\sqrt{2}} + \frac{i}{\sqrt{2}} \left( e^{i\phi_0} D[i\eta_x]D[i\eta_y]D[i\eta_z] \hat{\sigma}_+ + e^{-i\phi_0} D[-i\eta_x]D[-i\eta_y]D[-i\eta_z] \hat{\sigma}_- \right).
\label{eq:2}
\end{equation}
First, we apply $O_{N, \frac{\pi}{2}}$ to the initial state $\ket{\Psi}$.
\begin{eqnarray}
O_{N, \frac{\pi}{2}} \ket{\Psi} &=& \frac{1}{\sqrt{2}} + \frac{i}{\sqrt{2}} \left( e^{i\phi_0} D[i\eta_x]D[i\eta_y]D[i\eta_z] \hat{\sigma}_+ + e^{-i\phi_0} D[-i\eta_x]D[-i\eta_y]D[-i\eta_z] \hat{\sigma}_- \right) \ket{\downarrow, \alpha_x, \alpha_y, \alpha_z} \nonumber \\
&=& \frac{1}{\sqrt{2}} \ket{\downarrow, \alpha_x, \alpha_y, \alpha_z} + \frac{i}{\sqrt{2}} \left(e^{i\phi_0} \ket{\uparrow} e^{i\eta_x \mathrm{Re}[\alpha_x]} \ket{\alpha_x + i\eta_x} e^{i\eta_y \mathrm{Re}[\alpha_y]} \ket{\alpha_y + i\eta_y} e^{i\eta_z \mathrm{Re}[\alpha_z]} \ket{\alpha_z + i\eta_z} \right) \\
&&\equiv \ket{\Psi_1}. \nonumber
\end{eqnarray}
After the time $T_{\mathrm{int}}$, $\ket{\Psi_1}$ evolve in time,
\begin{eqnarray}
\ket{\Psi_1}_{T_{\mathrm{int}}} &=& \frac{1}{\sqrt{2}} \ket{\downarrow, \alpha_x e^{-i\theta_x}, \alpha_y e^{-i\theta_y}, \alpha_z e^{-i\theta_z}} \nonumber \\
&&+ \frac{i}{\sqrt{2}} \left(e^{i\phi_0} \ket{\uparrow} e^{i\eta_x \mathrm{Re}[\alpha_x]} \ket{(\alpha_x + i\eta_x) e^{-i\theta_x}} e^{i\eta_y \mathrm{Re}[\alpha_y]} \ket{(\alpha_y + i\eta_y)e^{-i\theta_y}} e^{i\eta_z \mathrm{Re}[\alpha_z]} \ket{(\alpha_z + i\eta_z) e^{-i\theta_z}} \right) \\
&\equiv& \ket{\Psi_2}, \nonumber 
\end{eqnarray}
where $\theta_j = \omega_j T_{\mathrm{int}} (j = x, y, z)$. Also apply the second $\pi/2$ pulse to $\ket{\Psi_2}$, we obtain 
\begin{eqnarray}
O_{N, \frac{\pi}{2}} \ket{\Psi_2} &=& \left\{ \frac{1}{\sqrt{2}} + \frac{i}{\sqrt{2}} \left( e^{i\phi_0} D[i\eta_x]D[i\eta_y]D[i\eta_z] \hat{\sigma}_+ 
+ e^{-i\phi_0} D[-i\eta_x]D[-i\eta_y]D[-i\eta_z] \hat{\sigma}_- \right) \right\} \nonumber \\
&&\times \left\{ \frac{1}{\sqrt{2}} \ket{\downarrow, \alpha_x e^{-i\theta_x}, \alpha_y e^{-i\theta_y}, \alpha_z e^{-i\theta_z}} + \frac{i}{\sqrt{2}} \left(e^{i\phi_0} \ket{\uparrow} e^{i\eta_x \mathrm{Re}[\alpha_x]} \ket{(\alpha_x + i\eta_x) e^{-i\theta_x}} \right. \right. \nonumber \\
&& \left. \left. \times e^{i\eta_y \mathrm{Re}[\alpha_y]} \ket{(\alpha_y + i\eta_y)e^{-i\theta_y}} e^{i\eta_z \mathrm{Re}[\alpha_z]} \ket{(\alpha_z + i\eta_z) e^{-i\theta_z}} \right) \right\} \nonumber \\
&=& \frac{1}{2} \ket{\downarrow, \alpha_x e^{-i\theta_x}, \alpha_y e^{-i\theta_y}, \alpha_z e^{-i\theta_z}} \nonumber \\
&&+ \frac{i}{2} e^{i\phi_0} \ket{\uparrow} e^{i\eta_x \mathrm{Re}[\alpha_x]}\ket{(\alpha_x + i\eta_x) e^{-i\theta_x}}
e^{i\eta_y \mathrm{Re}[\alpha_y]}\ket{(\alpha_y + i\eta_y) e^{-i\theta_y}} 
e^{i\eta_z \mathrm{Re}[\alpha_z]}\ket{(\alpha_z + i\eta_z) e^{-i\theta_z}} \nonumber \\
&&+\frac{i}{2} e^{i\phi_0} \ket{\uparrow} e^{i\eta_x \mathrm{Re}[\alpha_x e^{-i\theta_x}]} \ket{\alpha_x e^{-i\theta_x} + i\eta_x} e^{i\eta_y \mathrm{Re}[\alpha_y e^{-i\theta_y}]} \ket{\alpha_y e^{-i\theta_y} + i\eta_y} e^{i\eta_z \mathrm{Re}[\alpha_z e^{-i\theta_z}]} \ket{\alpha_z e^{-i\theta_z} + i\eta_z} \nonumber \\
&&-\frac{1}{2} \ket{\downarrow} e^{i\eta_x \mathrm{Re}[\alpha_x]} e^{-i\eta_x \mathrm{Re}[(\alpha_x + i\eta_x) e^{-i\theta_x}]} \ket{(\alpha_x + i\eta_x) e^{-i\theta_x} -i\eta_x} \nonumber \\
&&\times e^{i\eta_y \mathrm{Re}[\alpha_y]} e^{-i\eta_y \mathrm{Re}[(\alpha_y + i\eta_y) e^{-i\theta_y}]} \ket{(\alpha_y + i\eta_y) e^{-i\theta_y} -i\eta_y} \nonumber \\
&&\times e^{i\eta_z \mathrm{Re}[\alpha_z]} e^{-i\eta_z \mathrm{Re}[(\alpha_z + i\eta_z) e^{-i\theta_z}]} \ket{(\alpha_z + i\eta_z) e^{-i\theta_z} -i\eta_z} \\
&\equiv& \ket{\Psi_3}. \nonumber
\end{eqnarray}
The final measured brightness $B$ in Eq.~(5.20) in Ref.~[22] is defined as follow:
\begin{equation}
B = \int d^2\alpha P(\alpha) B(\alpha),
\label{bright}
\end{equation}
where $B(\alpha)$ is 
\begin{equation}
B(\alpha) = \braket{\Psi_3|\uparrow} \braket{\uparrow|\Psi_3}.
\end{equation}
We calculate $\braket{\Psi_3 | \uparrow}$ and $\braket{\uparrow | \Psi_3}$ respectively, 
\begin{eqnarray}
\braket{\Psi_3|\uparrow} &=& -\frac{i}{2} e^{-\phi_0} e^{-i\eta_x \mathrm{Re}[\alpha_x]} \bra{(\alpha_x + i\eta_x) e^{-i\theta_x}} e^{-i\eta_y \mathrm{Re}[\alpha_y]} \bra{(\alpha_y + i\eta_y) e^{-i\theta_y}} e^{-i\eta_z \mathrm{Re}[\alpha_z]} \bra{(\alpha_z + i\eta_z) e^{-i\theta_z}} \nonumber \\
&&-\frac{i}{2}e^{-i\phi_0} e^{i\eta_x \mathrm{Re}[\alpha_x]e^{-i\theta_x}} \bra{\alpha_x e^{-i\theta_x} + i\eta_x}
e^{i\eta_y \mathrm{Re}[\alpha_y]e^{-i\theta_y}} \bra{\alpha_y e^{-i\theta_y} + i\eta_y}
e^{i\eta_z \mathrm{Re}[\alpha_z]e^{-i\theta_z}} \bra{\alpha_z e^{-i\theta_z} + i\eta_z},
\end{eqnarray}
\begin{eqnarray}
\braket{\uparrow|\Psi_3} &=& \frac{i}{2}e^{i\phi_0} e^{i\eta_x \mathrm{Re}[\alpha_x]} \ket{(\alpha_x + i\eta_x) e^{-i\theta_x}}
e^{i\eta_y \mathrm{Re}[\alpha_y]} \ket{(\alpha_y + i\eta_y) e^{-i\theta_y}} 
e^{i\eta_z \mathrm{Re}[\alpha_z]} \ket{(\alpha_z + i\eta_z) e^{-i\theta_z}} \nonumber \\
&&+ \frac{i}{2}e^{i\phi_0} e^{i\eta_x \mathrm{Re}[\alpha_x e^{-i\theta_x}]} \ket{\alpha_x e^{-i\theta_x} + i\eta_x} e^{i\eta_y \mathrm{Re}[\alpha_y e^{-i\theta_y}]} \ket{\alpha_y e^{-i\theta_y} + i\eta_y} e^{i\eta_z \mathrm{Re}[\alpha_z e^{-i\theta_z}]} \ket{\alpha_z e^{-i\theta_z} + i\eta_z}.
\end{eqnarray}
Therefore, $B(\alpha)$ can be calculated
\begin{eqnarray}
B(\alpha) &=& \braket{\Psi_3|\uparrow}\braket{\uparrow|\Psi_3} \\
&=& \frac{1}{2} + \frac{1}{4} e^{-i\eta_x \mathrm{Re}[\alpha_x]}e^{i\eta_x \mathrm{Re}[\alpha_x]e^{-i\theta_x}}\braket{(\alpha_x + i\eta_x) e^{-i\theta_x} | \alpha_x e^{-i\theta_x} + i\eta_x} \nonumber \\
&&\times e^{-i\eta_y \mathrm{Re}[\alpha_y]}e^{i\eta_y \mathrm{Re}[\alpha_y]e^{-i\theta_y}}\braket{(\alpha_y + i\eta_y) e^{-i\theta_y} | \alpha_y e^{-i\theta_y} + i\eta_y} \nonumber \\
&&\times  e^{-i\eta_z \mathrm{Re}[\alpha_z]}e^{i\eta_z \mathrm{Re}[\alpha_z]e^{-i\theta_z}}\braket{(\alpha_z + i\eta_z) e^{-i\theta_z} | \alpha_z e^{-i\theta_z} + i\eta_z} \nonumber \\
&&+ \frac{1}{4} e^{-i\eta_x \mathrm{Re}[\alpha_x e^{-i\theta_x}]} e^{i\eta_x \mathrm{Re}[\alpha_x]} \braket{\alpha_x e^{-i\theta_x} + i\eta_x | (\alpha_x + i\eta_x)e^{-i\theta_x}} \nonumber \\
&&\times e^{-i\eta_y \mathrm{Re}[\alpha_y e^{-i\theta_y}]} e^{i\eta_y \mathrm{Re}[\alpha_y]} \braket{\alpha_y e^{-i\theta_y} + i\eta_y | (\alpha_y + i\eta_y)e^{-i\theta_y}} \nonumber \\
&&\times e^{-i\eta_z \mathrm{Re}[\alpha_z e^{-i\theta_z}]} e^{i\eta_z \mathrm{Re}[\alpha_z]} \braket{\alpha_z e^{-i\theta_z} + i\eta_z | (\alpha_z + i\eta_z)e^{-i\theta_z}} \nonumber \\
&=& \frac{1}{2} + \frac{1}{4} \left(e^{-(\eta_x^2 - \eta_x^2 e^{i\theta_x} + 2i\gamma_x)}
e^{-(\eta_y^2 - \eta_y^2 e^{i\theta_y} + 2i\gamma_y)} 
e^{-(\eta_z^2 - \eta_z^2 e^{i\theta_z} + 2i\gamma_z)} \right. \nonumber \\
&&\left. + e^{-(\eta_x^2 - \eta_x^2 e^{-i\theta_x} - 2i\gamma_x)}
e^{-(\eta_y^2 - \eta_y^2 e^{-i\theta_y} - 2i\gamma_y)}
e^{-(\eta_z^2 - \eta_z^2 e^{-i\theta_z} - 2i\gamma_z)} \right) \\
&=& \frac{1}{2} \left( 1 + e^{-\eta_x (1-\cos \theta_x)} \cos (\eta_x^2 \sin \theta_x - 2\gamma_x) e^{-\eta_y (1-\cos \theta_y)} \right. \nonumber \\
&&\left. \times \cos (\eta_y^2 \sin \theta_y - 2\gamma_y) 
e^{-\eta_z (1-\cos \theta_z)} \cos (\eta_z^2 \sin \theta_z - 2\gamma_z) \right).
\end{eqnarray}
Here let $\alpha_j = a_j + ib_j (j = x, y, z)$, we obtain $\mathrm{Re}[\alpha_j] = a_j$, $\mathrm{Re}[\alpha_j e^{-i\theta_j}] = a_j \cos \theta_j + b_j \sin \theta_j$, thus we can let $\gamma_j$ as $\gamma_j = \eta_j (a_j - a_j \cos \theta_j - b_j \sin \theta_j)$.  We substitute them in Eq.~(\ref{bright}). Also, $P(\alpha) = \frac{1}{\pi \overline{n}} e^{-\frac{|\alpha|^2}{\overline{n}}}$ as shown in Eq.~(5.11) in Ref.~[22], while we consider in three dimension, we let $P(\alpha) = P(\alpha_x)P(\alpha_y)P(\alpha_z) = \frac{1}{\pi \overline{n}_x} e^{-\frac{|\alpha_x|^2}{\overline{n}_x}} 
\frac{1}{\pi \overline{n}_y} e^{-\frac{|\alpha_y|^2}{\overline{n}_y}} 
\frac{1}{\pi \overline{n}_z} e^{-\frac{|\alpha_z|^2}{\overline{n}_z}} $. We also substitute it in Eq.~(\ref{bright}):
\begin{eqnarray}
B &=& \int da_x db_x \frac{1}{\pi \overline{n}_x}e^{-\frac{1}{\overline{n}_x}(a_x^2 + b_x^2)}
\int da_y db_y \frac{1}{\pi \overline{n}_y}e^{-\frac{1}{\overline{n}_y}(a_y^2 + b_y^2)}
\int da_z db_z \frac{1}{\pi \overline{n}_z}e^{-\frac{1}{\overline{n}_z}(a_z^2 + b_z^2)}
\nonumber \\
&&\times \frac{1}{2} \left(1+ e^{-\eta_x^2 (1-\cos\theta_x)}\cos (\eta_x^2 \sin \theta_x - 2\gamma_x) e^{-\eta_y^2 (1-\cos\theta_y)}\cos (\eta_y^2 \sin \theta_y - 2\gamma_y) \right. \nonumber \\
&& \left. e^{-\eta_z^2 (1-\cos\theta_z)}\cos (\eta_z^2 \sin \theta_z - 2\gamma_z) \right) \\
&=& \frac{1}{2} \left[1+ e^{-\eta_x^2 (1+2\overline{n}_x)(1-\cos \theta_x)}\cos (\eta_x^2 \sin \theta_x) 
e^{-\eta_y^2 (1+2\overline{n}_y)(1-\cos \theta_y)}\cos (\eta_y^2 \sin \theta_y) \right. \nonumber \\
&& \left.e^{-\eta_z^2 (1+2\overline{n}_z)(1-\cos \theta_z)}\cos (\eta_z^2 \sin \theta_z) \right].
\end{eqnarray}
Consequently, this is the equation of our interferometry in the manuscript.